\documentclass[journal]{IEEEtran}
\usepackage{booktabs}
\usepackage{lineno,hyperref}
\usepackage[english]{babel}
\usepackage{amsmath}
\usepackage{bm}
\usepackage{graphicx}
\usepackage{subfigure}
\usepackage{tabularx}
\usepackage{multirow}
\usepackage{multicol}
\usepackage{caption}
\usepackage{float}
\usepackage{cite}
\usepackage{epstopdf}
\usepackage{color}
\usepackage{algpseudocode}
\usepackage[ruled,linesnumbered]{algorithm2e}
\usepackage{enumerate}
\usepackage{amsthm, amsfonts}
\newtheorem{myDef}{Definition}

\modulolinenumbers[5]

\ifCLASSINFOpdf
\else
\fi

\begin{document}
\captionsetup[figure]{name={Fig.}}
% \title{Time-Series Snapshot Network as A New Model for Role Recommendation in OSS}
% \title{Time-Series Snapshot Network as A New Model for Link Prediction in OSS}
\title{Time-Series Snapshot Network for Partner Recommendation: A Case Study on OSS}

\author{Yunyi Xie, 
Jinyin Chen,
Jian Zhang,
Xincheng Shu, and 
Qi Xuan,~\IEEEmembership{Member,~IEEE} % <-this % stops a space
% \thanks{This work was partially supported by National Natural Science Foundation of China (61572439), Zhejiang Provincial Natural Science Foundation of China (LR19F030001, LY19F020025), the Major Special Funding for “Science and Technology Innovation 2025” in Ningbo (2018B10063), the National Key Research and Development Program of China (2018AAA0100800). \emph{(Corresponding author: Qi Xuan.)}}
\thanks{Y. Xie, J. Chen, J. Zhang, X Shu,and Q. Xuan are with the Institute of Cyberspace Security and the College of Information Engineering, Zhejiang University of Technology, Hangzhou 310023, China (e-mail: yunyixie97@foxmail.com; chenjinyin@zjut.edu.cn; jianzh@zjut.edu.cn; sxc.shuxincheng@foxmail.com; xuanqi@zjut.edu.cn).}
\thanks{Y. Xie and J. Chen contributed equally to this work and should be joint first authors.}
\thanks{Corresponding author: Qi Xuan.}
}

%to creep in.
% IEEE Transactions on Computational Social Systems
% IEEE Transactions on Network Science and Engineering
% \markboth{IEEE Transactions on Computational Social Systems}%
% {Xuan \MakeLowercase{\textit{et al.}}: Human foraging patterns}
\maketitle

\begin{abstract}
The last decade has witnessed the rapid growth of open source software~(OSS). Still, all contributors may find it difficult to assimilate into OSS community even they are enthusiastic to make contributions. We thus suggest that partner recommendation across different roles may benefit both the users and developers, i.e., once we are able to make successful recommendation for those in need, it may dramatically contribute to the productivity of developers and the enthusiasm of users, thus further boosting OSS projects' development. Motivated by this potential, we model the partner recommendation as link prediction task from email data via network embedding methods. In this paper, we introduce time-series snapshot network~(TSSN) which is a mixture network to model the interactions among users and developers. Based on the established TSSN, we perform temporal biased walk~(TBW) to automatically capture both temporal and structural information of the email network, i.e., the behavioral similarity between individuals in the OSS email network. Experiments on ten Apache datasets demonstrate that the proposed TBW significantly outperforms a number of advanced random walk based embedding methods, leading to the state-of-the-art recommendation performance. 
\end{abstract}


\IEEEpeerreviewmaketitle

% \section{Introduction}\label{Section1}
% % The very first letter is a 2 line initial drop letter followed
% % by the rest of the first word in caps.
% % 
% % form to use if the first word consists of a single letter:
% % \IEEEPARstart{A}{demo} file is ....
% % 
% % form to use if you need the single drop letter followed by
% % normal text (unknown if ever used by the IEEE):
% % \IEEEPARstart{A}{}demo file is ....
% % 
% % Some journals put the first two words in caps:
% % \IEEEPARstart{T}{his demo} file is ....
% % 
% % Here we have the typical use of a "T" for an initial drop letter
% % and "HIS" in caps to complete the first word.

\section{Introduction}\label{sec:introduction}
\IEEEPARstart{T}{he} rapid growth of open source software~(OSS) has risen to great prominence within the last several years. A large number of users are attracted to join the OSS community~\cite{fronchetti2019attracts}. Developers and users' active engagement is crucial to the success of OSS projects~\cite{steinmacher2014hard}. To promote OSS projects' sustainable development, it is necessary for developers to maintain projects' code~\cite{xia2015should},~\cite{yu2014should},~\cite{fejzer2018profile}. Similarly, it is essential to motivate, engage, and retain users and developers~\cite{steinmacher2014attracting}.

The participation of users and developers in OSS projects requires to overcome many obstacles, which discourages further contributions to OSS projects~\cite{steinmacher2015systematic},~\cite{steinmacher2014preliminary}. As the mailing list is public communication channels in the OSS community, users and developers often use such means to start their interactions in the projects~\cite{steinmacher2013newcomers}, i.e., those who lack understanding and guidance generally post their questions and request help or exploit existing information available in the mailing list. As an example, a user named Shaw solved the doubts about "Airflow" project by posting questions on the mailing list to ask for help and he received numerous replies to solve his doubts~\cite{emaillist}: the question \emph{Global Custom Template Variables? Or is there another best practise here?} and one of the replies \emph{Re: Global Custom Template Variables? Or is there another best practise here?}. However, it's not easy to access information due to the large volume and useless replies. The barriers that users and developers face will lead them to give up their further contributions to OSS. Given this scenario, it may be helpful to recommend experienced developers and users to those who need help so that they can send emails to specific individuals to avoid more frustration. However, it's difficult to identify right and/or relevant people to help those in need. Identifying the right people is an important task and we often use a linguistic perspective to solve this challenge~\cite{liao2019status}. 

Partner recommendation in OSS can benefit both the users and developers. For instance, after recommending relevant developers for those who need assistance, they can send emails to someone who is relevant for help rather than search for needed in the information explosion mailing list. Similarly, if we recommend users to developers, the latter can actively provide help to the former in need and thus improve the participation of developers in projects. In this paper, we focus on the study of recommending developers and users to each other in the OSS community from network perspective and model partner recommendation as link prediction task in OSS social networks. It could be useful to model the mailing list using a graph-based approach rather than a general method since the mailing list data is publicly available and universally applicable in social network analysis. 

% The proposed method relies on the network structure of the mail corpus composed of the OSS mailing list, where users and developers are vertices in the email network and temporal edges capture email exchanges. To further study the OSS email networks, we define a mixture network namely time-series snapshot network~(TSSN). Based on TSSN, we further introduce a temporal biased walk~(TBW) to learn the representations of users and developers. For each user/developer, an unique searching strategy is adopted. The searching strategy depends on the number of email exchanges, structure-based transition probability, temporal transition probability, and role-based transition probability. Our approach efficiently learns individuals' representations based on the proposed TBW and makes role recommendation via machine learning techniques. Furthermore, we present some preliminary analysis of OSS email datasets which motivated the design of our method and verify the necessity to make role recommendation in OSS. 

We construct the email network that relies on the network structure of the mail corpus composed of the OSS mailing list to gain better insight into the OSS mailing list, where users and developers are vertices in the email network and temporal edges capture email exchanges. To further retain the email's temporal property, we define a mixture network namely time-series snapshot network~(TSSN). Based on TSSN, we further introduce a temporal biased walk~(TBW) to learn the representations of users and developers. We believe that, by comparison, the embeddedness of developers and users' behavior on graph can capture more important interactions of individuals. For each user/developer, an unique searching strategy is adopted. The searching strategy depends on the number of email exchanges, structure-based transition probability, temporal transition probability, and role-based transition probability. Our approach efficiently learns individuals' representations based on the proposed TBW and makes partner recommendation via machine learning techniques. Also, we present some preliminary analysis of OSS email datasets which motivated the design of our method and verify the necessity to make partner recommendation across different roles in OSS. 

The main contributions of this paper are summarized as follows. 
\begin{itemize}
\item We study the matter of recommending partners for individuals who need assistance in the OSS community from a network perspective. When individuals encounter difficulties, such recommendations can provide them with certain support, which is crucial for OSS projects' sustainable development. In addition, the data used in this paper has been cleaned up and standardized and will be available online for future study.
\item We construct time-series snapshot network~(TSSN) for OSS mailing list to capture the evolution of the interactions among users and developers, and thus retain OSS email's temporal property. Moreover, we propose a temporal biased walk~(TBW) to effectively embed developers and users based on their interactions, which integrates temporal information, structural information, and individuals' identities of the OSS email networks. Furthermore, our method is more applicable and can be modified for other datasets on their unique properties.
%Considering the OSS mailing list can be modeled as a temporal network, we define a mixture network named TSSN to express the evolution of network. On this basis, we further propose a graph-based random walk TSSNW. It's an effective method that combines temporal information, network structure information, and roles' real identity on the OSS email networks.
\item We carry on several types of recommendation experiments on realistic OSS projects, and the results demonstrate that our proposed TBW significantly outperforms a number of random walk based embedding methods in partner recommendation. More precisely, our method combined with OSS dataset's properties can be more superior to other general machine learning methods. 

%Besides, we gain insights by putting TSSNW into practice that our label biased sampling can indeed obtain a variety of samplings and thus enhance recommendation results.
\end{itemize}

The remainder of this paper is organized as follows. In Sec.~\ref{sec:relatedwork}, we review the related works in OSS email network analysis and random walk based embedding methods. Then, we present the data collection and preliminary analysis of OSS datasets in Sec.~\ref{sec:ossdataset}. After that, in Sec.~\ref{sec:theproposedmethods} we give the basic definition and our proposed method, and in Sec.~\ref{sec:experiments} we conduct extensive experiments with discussions. In Sec.~\ref{sec:Threat}, we give the possible threats to validity. Finally, we conclude the paper and highlight future work in Sec.~\ref{sec:Conclusion}. 

\section{Related work} \label{sec:relatedwork}
% In this section, we review some OSS network analysis and network embedding algorithms based on random walk methods in recent years.
\subsection{OSS Analysis}
In OSS projects, the interaction between developers and users, as well as the code submission, is usually public and stored for future reference. Therefore, the project archive provides rich historical information resources that can be used to study many fascinating matters such as discovering potential cooperation between users and developers. 

Most researchers usually use the OSS mailing list data for quantitative analysis to gain insight into the social aspects of software development and provide relevant insights~\cite{guzzi2013communication},~\cite{shihab2009central}. For example, Bird et al. proposed the technology of mining OSS email networks~\cite{bird2006mining}, and presented some preliminary results from email network analysis. In~\cite{bird2008latent}, they studied the social interaction in OSS projects and discovered the latent social structure. Xuan et al.~\cite{xuan2012measuring} proposed a novel quantitative method to measure the impact of social communication on individual work rhythm by analyzing communication and code submission records in OSS projects. The results showed that mailing list activity is strongly related to source code activity. More recently, they proposed another quantitative method~\cite{xuan2014building} to identify synchronization activities in OSS projects and use them to connect developer synchronization with effective productivity and communication. Most of the aforementioned work is quantitative and based on the mailing list communication to study the relationship between users and developers. This is mainly because the interaction between developers and users is crucial to OSS projects' development.

Maintaining projects' code and retaining users to actively participate in the projects are equally essential for OSS projects' sustainable development. However, most researches focus on selecting excellent developers to maintain the stability of the projects' code, while ignoring the importance of users in the projects. For instance, Lee et al.~\cite{lee2013patch} proposed a graph-based method to automatically recommend experienced developers to review patches before applying or submitting them. Ouni et al.~\cite{ouni2016search} introduced a search-based approach to find the most appropriate reviewers for submitted code changes for improving software quality and reducing defect proneness. Kagdi and Poshyvanyk recommended a ranked list of developers to assist in performing software changes~\cite{kagdi2009can}. Fu et al. designed a recommender system to recommend appropriate experts for developers to integrate new developers with team~\cite{fu2017expert}. We argue that such preference may promote the short-term code contribution but may discourage the users and thus hurt the long-term development of the OSS community.  

%Few studies focus on OSS social networks to explore the interaction between developers and users who need support. 
Therefore, in this paper, we address users and developers equally and try to support them in real time by proposing a TBW based partner recommendation algorithm in TSSN, which considers users and developers as two kinds of roles. 
%Users and developers will encounter various difficulties when participating in OSS, thus it is necessary to recommend experienced individuals to help them, e.g., give suggestions or assist in commits. In this work, we explore users and developers' behavior in terms of embeddedness in social networks via social perspective. We conduct the preliminary analysis of developers and users based on the OSS mailing list and recommend developers as well as users for individuals who need help to support them. In this way, seeking help from specific individuals can indeed promote OSS projects' sustainable development.

\subsection{Random Walk Based Network Embedding}
Network embedding has received much attention over the past decades~\cite{cui2018survey},~\cite{hamilton2017representation}. Such methods intuitively focus on transforming each vertex into a low-dimensional vector based on its local structure in the network. Since vertices that share similar structural properties are closed to each other in the low-dimensional embedding space, one can easily use the learned embeddings for downstream tasks such as community detection~\cite{fortunato2010community}, link prediction~\cite{lu2011link}, and node classification~\cite{rossi2012time}. 

One of the earliest efforts in network embedding is to combine random walk based methods with Skip-Gram~\cite{mikolov2013efficient} model to learn vertex representation, where Skip-Gram model was first introduced in the natural language processing~(NLP) domain~\cite{rong2014word2vec}. The theory underlying random walks and their connection to eigenvalues and other fundamental properties of graphs are well-understood~\cite{chung2007random},~\cite{mohan2019network}. Thus, various random walk based embedding methods~\cite{perozzi2014deepwalk},~\cite{tang2015line},~\cite{grover2016node2vec}, combined with Skip-Gram model, have been vastly proposed. 

Two of the most notable random walk based methods are DeepWalk~\cite{perozzi2014deepwalk} and Node2vec~\cite{grover2016node2vec}. DeepWalk is one of the earliest work in network embedding, which uses the random walk to generate sequences for each vertex. The sequences of vertices are treated as text in language models, based on which one can learn vertex representation by Skip-Gram model. Node2vec~\cite{grover2016node2vec} is a biased second-order random walk model that extends DeepWalk by employing biased random walks to learn vertex embedding, and it can capture both vertex homophily and structural equivalence. Tang et al.~\cite{tang2015line} proposed another successful network embedding model LINE, which designs the objective function, optimizes the first-order and second-order proximity, and performs the optimizations by stochastic gradient descent with edge sampling. BiasedWalk~\cite{nguyen2018biasedwalk} overcomes the disadvantage that Node2vec needs to store the interactive information of vertices. It adopts simulation biased random walk to balance the breadth-first and depth-first graph traversal.

With the development of random walk based methods in network embedding, more and more researchers have tried to apply random walk on temporal networks~\cite{acer2010random}. Dynnode2vec~\cite{mahdavi2018dynnode2vec} modifies Node2vec by considering the previous embedding vectors as the initialization of Skip-Gram model, and employing random walks in network evolution to update Skip-Gram model based on previous timestamps. However, they didn't consider the directly correlation between different snapshots, which may lead to extra loss. Nguyen et al.~\cite{nguyen2018continuous},~\cite{lee2019temporal} proposed a general framework CTDNE. It's a new class of embeddings learned directly from the temporal network~(graph stream) without having to approximate the edge stream as a sequence of discrete static snapshot graphs. However, it does not consider the edge's weight and may cause incomplete representations.

Considering the high expressiveness and learning ability of random walk based methods with different search strategies, we model general temporal networks as a spatial-temporal network, i.e., time-series snapshot network~(TSSN). Furthermore, we implement a random walk using specific search strategies, namely temporal biased walk~(TBW), on the proposed TSSN. In this algorithm, each vertex has its unique search strategy, reflecting its global and local structural properties in spatial and temporal domains simultaneously. 

\section{OSS dataset}\label{sec:ossdataset}
In this section, we first give a detailed description of the dataset and data preprocessing, and then conduct preliminary descriptive analyses that motivate our methodology in Sec.~\ref{sec:theproposedmethods} and indicate the reason to make partner recommendation across different roles in OSS. 

\subsection{Data Description \& Preprocessing}
The Apache Software Foundation~(ASF) is the largest open source foundation in the world and incubates hundreds of free enterprise projects, such as \emph{Hadoop} and \emph{ApacheHTTP}, which act as the backbone of some widely used applications. ASF opens not only source codes but also emails among individuals of all projects, making the ASF project archive a rare public data source for social and technical activity analysis. We gather email data by parsing the email activities on the Apache mailing list over a period starting from 1999 to Aug 2019. Besides, 10 projects which graduate from the incubator are selected as the representatives and we conduct further study according to the projects' mailing list. Here, we focus on the projects graduated from the incubator. In particular, we first choose 20 projects with the maximum number of developers, among which we then select 10 most active ones, i.e., with most email communications. 

%we choose 10 projects from hundreds of projects. These 10 projects are the top 20 projects in the number of developers out of hundreds of projects, and then the top 10 projects based on the number of emails are selected.

% \textcolor{red}{Besides, 10 projects which graduate from the incubator are selected as the representatives and we conduct further study according to the projects' mailing list. The proportion of users and developers in the project mailing list and the number of email exchanges are the basis for selecting these projects.} 

We first extract recipient, sender, sending time, title and content of each email record contained in the mailing list before constructing network. To accurately scale an individual's activity, we introduce identity matching to merge different aliases. 
% The proposed dealiasing algorithm might have lower performance when it comes to the aliases associated with little activity. Alternative dealiasing methods like Christian Bird~\cite{bird2006mining} and Igor Scaliante \cite{wiese2016mailing} probably are more accurate but less efficient to deal with tens of thousands of aliases. 
In subsequent experiments, we retain only the date stamp of each email, anonymized sender and recipient IDs. Based on a number of projects' emails extracted from the Apache mailing list, we construct several email networks. In every network, each vertex denotes either a user or a developer, and the weighted edge represents the email exchanges. Note that those with the same sender and recipient are removed for that such emails cannot be regarded as social interactions. The summary of the 10 projects and their descriptive statistics are provided in Table~\ref{tab:ossdata}.
\begin{table}[htbp]
  \centering
  \renewcommand{\arraystretch}{1.2}
  \setlength{\abovecaptionskip}{0pt}
  \setlength{\belowcaptionskip}{5pt}
  \caption{Overview of Apache datasets.}
  \begin{tabular}{cccccc}
  \toprule
  Project    & Users & Developers & Email exchanges & \begin{tabular}[c]{@{}c@{}}Timespan\\ (month)\end{tabular}                                                        \\ \midrule
  Airflow    & 411   & 95         & 4493            & 33               \\
  Beam       & 159   & 61         & 3363            & 11               \\
  CarbonData & 123   & 53         & 1477            & 11               \\
  Cordova    & 622   & 61         & 9832            & 14               \\
  Geode      & 179   & 60         & 5860            & 21               \\
  HAWQ       & 169   & 62         & 4015            & 36               \\
  Impala     & 152   & 39         & 3511            & 25               \\
  Metron     & 109   & 48         & 5445            & 17               \\
  Spark      & 223   & 45         & 2707            & 10               \\
  Zeppelin   & 379   & 67         & 5924            & 18               \\ \bottomrule
  \end{tabular}
  \label{tab:ossdata}
\end{table}

\subsection{Preliminary Analysis}\label{sec:preliminaryanalysis}
\label{sec:roles}
In each Apache project, we classify all projects' participants into one of the following categories~\cite{kamei2008analysis}:
\begin{itemize}
  \item \textbf{Users} are individuals who use the software. They provide feedback to developers in the form of bug reports and feature suggestions to contribute to the Apache projects, and help other users join the Apache community through mailing lists and support forums.
  \item \textbf{Developers} actively participate in the project and contribute to code and documentation. They are also active in the mailing list, participate in discussions, provide patches, documentation, suggestions, and criticism. 
\end{itemize}
These roles can change over time, e.g., in July 2019, John may have been only reviewing patches or asking questions, but in August 2020, he may have submitted code changes/patches. 
% In this paper, we suggest that if a user has submitted code, he will be a developer throughout the development of the project.
In this paper, we suggest that once an individual has code submission behavior in a project, he/she will be treated as a developer of this project. Those who only send and receive emails are considered users. For different projects, the same individual may play different roles, i.e., a user of a project but a developer of another.

\begin{table}[htbp]
  \centering
  \renewcommand{\arraystretch}{1.2}
  \setlength{\abovecaptionskip}{0pt}
  \setlength{\belowcaptionskip}{5pt}
  \caption{T-test for the differences between users and developer in terms of emails received and sent.}
  \begin{tabular}{ccccc}
  \toprule
  %            & User      & Developer & T-value  & Significance\\ \midrule
  % Emails received  & 0.0024    & 0.0094    & 9.8611   & $p<0.001$   \\
  % Emails sent & 0.0028    & 0.0109    & 8.3245   & $p<0.001$    \\ \bottomrule
                       & User       & Developer & T-value  & Significance \\ \midrule
  \# Emails received   & 10.9596    & 44.4396    & 8.6787   & $p<0.001$   \\
  \# Emails sent       & 12.0478    & 54.0438    & 7.5854   & $p<0.001$    \\ \bottomrule

  \end{tabular}
  \label{tab:t-test}
\end{table}

Before introducing our method, we first take a glance at Apache dataset. To study the different roles of users and developers, we count the emails received and sent by users and developers, respectively, and then carry out a simple T-test as shown in Table~\ref{tab:t-test}. As we can see, there are statistically significant structural differences between users and developers in the email network. Therefore, we should consider the differences in email behavior between users and developers in the design of partner recommendation algorithm.
% Therefore, we should treat users and developers differently in the design of role recommendation algorithm.

\begin{figure}[htbp]
  \centering
  \includegraphics[width=\linewidth]{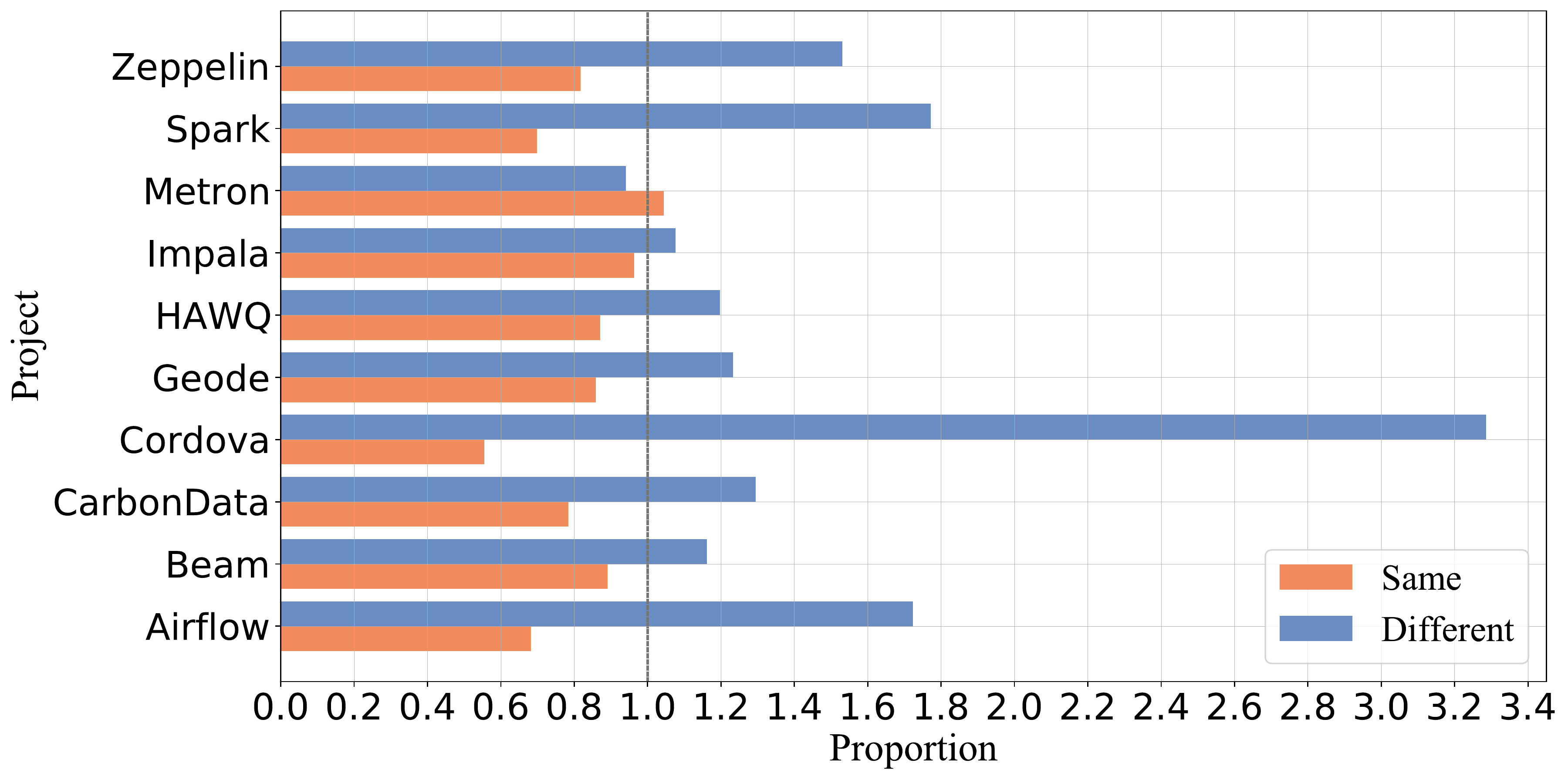}
  \caption{The comparison of real proportion and random proportion demonstrates communication tendency.}
  \label{fig:labeltrend}
\end{figure}

We further study the communication tendency of users and developers to see whether they are more likely to contact the individuals of the same roles or not. Here, we compare the proportion of same and different type roles' contact information with the random proportion of same and different information. If result $>1$, it reflects roles' real communication tendency. As shown in Fig.~\ref{fig:labeltrend}, we confirm that individuals tend to communicate with different ones. The results are not surprising because users are more likely to seek developer for help. Therefore, it's reasonable to consider users and developers' real identities in the design of partner recommendation algorithm.

%are not surprising because in Apache projects, individuals in the same type always have common topics and users account for a large proportion of participants in the projects. With these observations, we propose label-based sampling in Sec. \ref{sec:temporalwalk} to explore the whether participants' real identities will influence recommendation.
% \begin{figure}[htbp]
%   \centering
%   \includegraphics[width=\linewidth]{image/email_percent.pdf}
%   \caption{Email exchanges in random recommendation.}
%   \label{fig:emailpercent}
% \end{figure}

% We turn to a more qualitative study and thus we plot the proportion of email exchanges between these individuals who are successfully recommended in random recommendation in Figure \ref{fig:emailpercent}. Successfully recommended individuals refer to 25\% of those who have connections we randomly selected 10 times in each project. Interestingly, we find that a pair of individuals who are successfully recommended have more than 2 email exchanges, and the correspondence with more than 3 email exchanges also accounts for a large proportion. The results indicate that after being recommended, individuals can send emails to specific individuals and get replies, and they are likely to keep in touch for a long time. Therefore, we have reason to believe that role recommendation in OSS is very necessary, which can be used to make role recommendation to those in need.

\section{the proposed method}\label{sec:theproposedmethods}
In this section, we give several basic definitions, and then present the search strategy based on the spatial and temporal features of each user~(developer) in TSSN. The notations in this paper are listed in Table~\ref{tab:notations}.

\begin{table}[htbp]
  \centering
  \renewcommand{\arraystretch}{1.2}
  \setlength{\abovecaptionskip}{0pt}
  \setlength{\belowcaptionskip}{5pt}
  \caption{Notations used in this paper.}
  \begin{tabular}{lr}
  \toprule
    Symbol          & Description       \\ \midrule
    $G$             & The target graph~(network)  \\
    $V, E$          & Sets of vertices, edges in graph $G$  \\
    $\epsilon$                    & Time interval  \\
    $t \in \{0, 1, 2, \cdots\}$   & Time order     \\
    $G_{t}$         & The snapshot of graph $G$      \\
    $V_{t}, E_{t}$  & Sets of vertices, edges in graph $G_{t}$ \\
    $Src(e)$        & The source vertex of edge $e$     \\
    $Dst(e)$        & The target vertex of edge $e$      \\
    $W(e)$          & The weight of edge $e$             \\
    $T(e)$          & The time accessibility of edge $e$  \\
    $\eta_{+} : \mathbb{R} \to \mathbb{Z}^{+}$   & Function that maps vertex to index \\
    $L_{t}(v)$      & Set of accessible edges for vertex $v$  \\ \midrule
    $P(e)$          & The selection probability of edge $e$         \\
    $q$             & The in-out parameter          \\
    $r$             & The return parameter           \\
    $d_{tx}$        & The shortest path distance between vertex $t$ and $x$  \\     
    $\alpha$        & Temporal bias           \\
    $\beta$         & Role bias               \\
    $\omega(t)$     & The real identity of vertex $v$   \\ \midrule
    $N_{s}(v)$      & Set of temporal neighbors \\
    $f(v)$          & Mapping vertex $v$ to $d$-dimensional representation  \\ \bottomrule
  \end{tabular}
  \label{tab:notations}
\end{table}

\subsection{Basic Definition}\label{sec:definition}
\begin{figure*}[htbp]
  \centering
  \includegraphics[width=\linewidth]{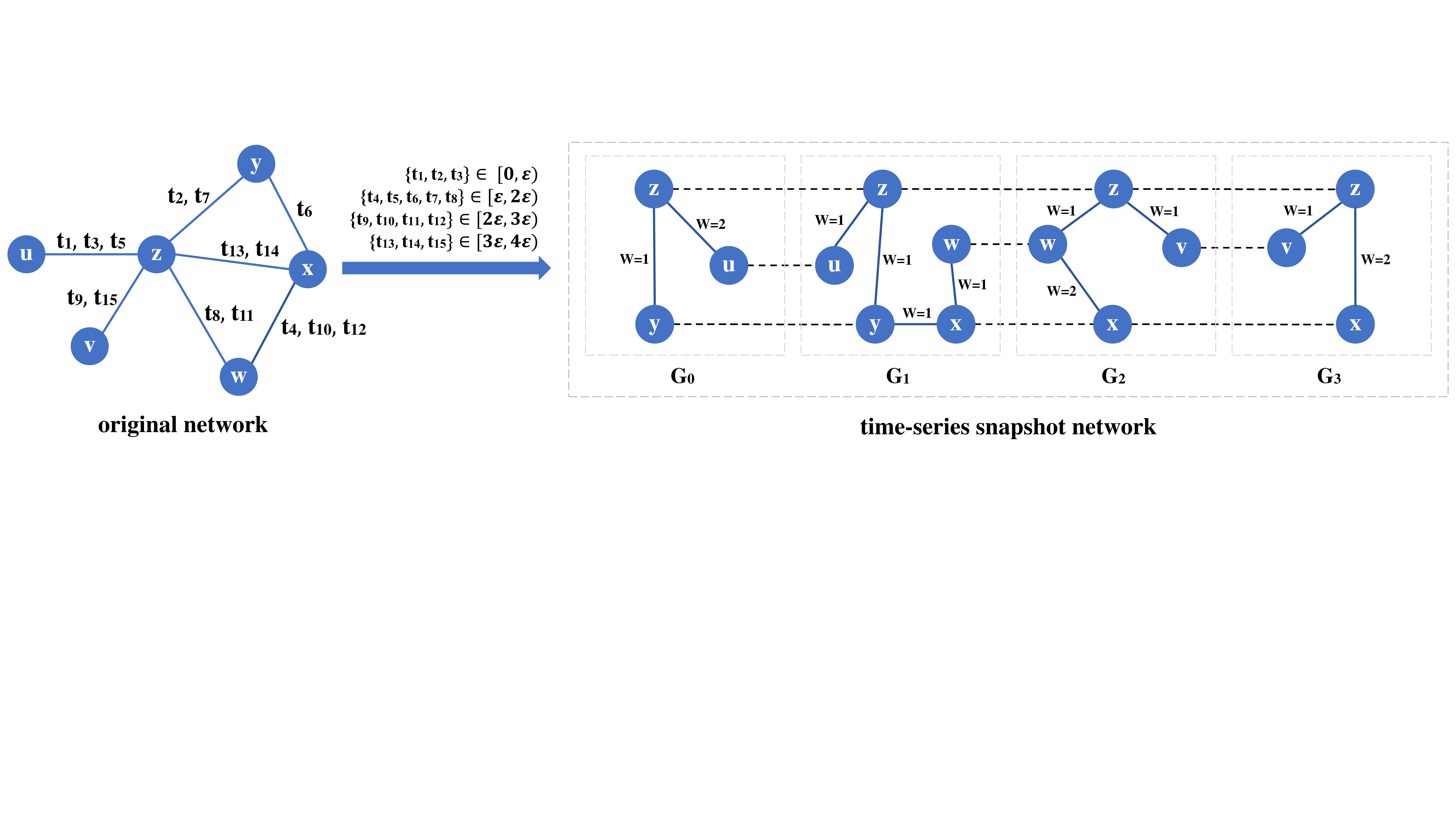}\caption{The detailed construction of time-series snapshot network~(TSSN). Dashed lines represent the self-connections, and solid lines denote the connection of pairwise vertices in the same snapshot.}
  \label{fig:process}
  \end{figure*}

In general, a project's mailing list can be modeled as a graph $G=(V,E)$ comprised of a vertex set $V$ with two types of vertices~(users and developers), and an edge set $E$ that represents the email exchanges with timestamps. On the basis of the given time interval $\epsilon$, we can divide the entire graph into several different snapshots. In order to capture the structural changing tendency of vertices in a temporal network, it is crucial to consider not only the snapshot at the current time, but also the nearby snapshots in time. %Utilizing the information at each snapshot separately will fail to capture the correlation information that exists between two graphs at consecutive time steps. 
Hence, we define TSSN, a spatial-temporal network shown in Fig.~\ref{fig:process}, to formulate our solution and further propose TBW to better capture the spatial and temporal properties of each vertex in TSSN, to facilitate the design of following partner recommendation. 

\begin{myDef}[\textbf{Time-Series Snapshot Network (TSSN)}] Given a graph $G=(V,E)$, which is divided into several snapshots $\{G_{0}, G_{1}, G_{2}, \cdots\}$ according to time span $\epsilon$, where $G_{t}=(V_{t},E_{t})$. Let $V_{t}$ and $E_{t}$ be sets of vertices and edges of snapshot $G_{t}$ respectively, in the timespan $[t\epsilon,(t+1)\epsilon)$, with time order $t \in \{0, 1, 2, \cdots\}$. All snapshots are sorted by time order $t$~(ascending). It's worth noting that self-connections could be established when and only when a node existed in successive snapshots. 
\end{myDef}

\begin{figure}[htbp]
  \centering
  \includegraphics[width=\linewidth]{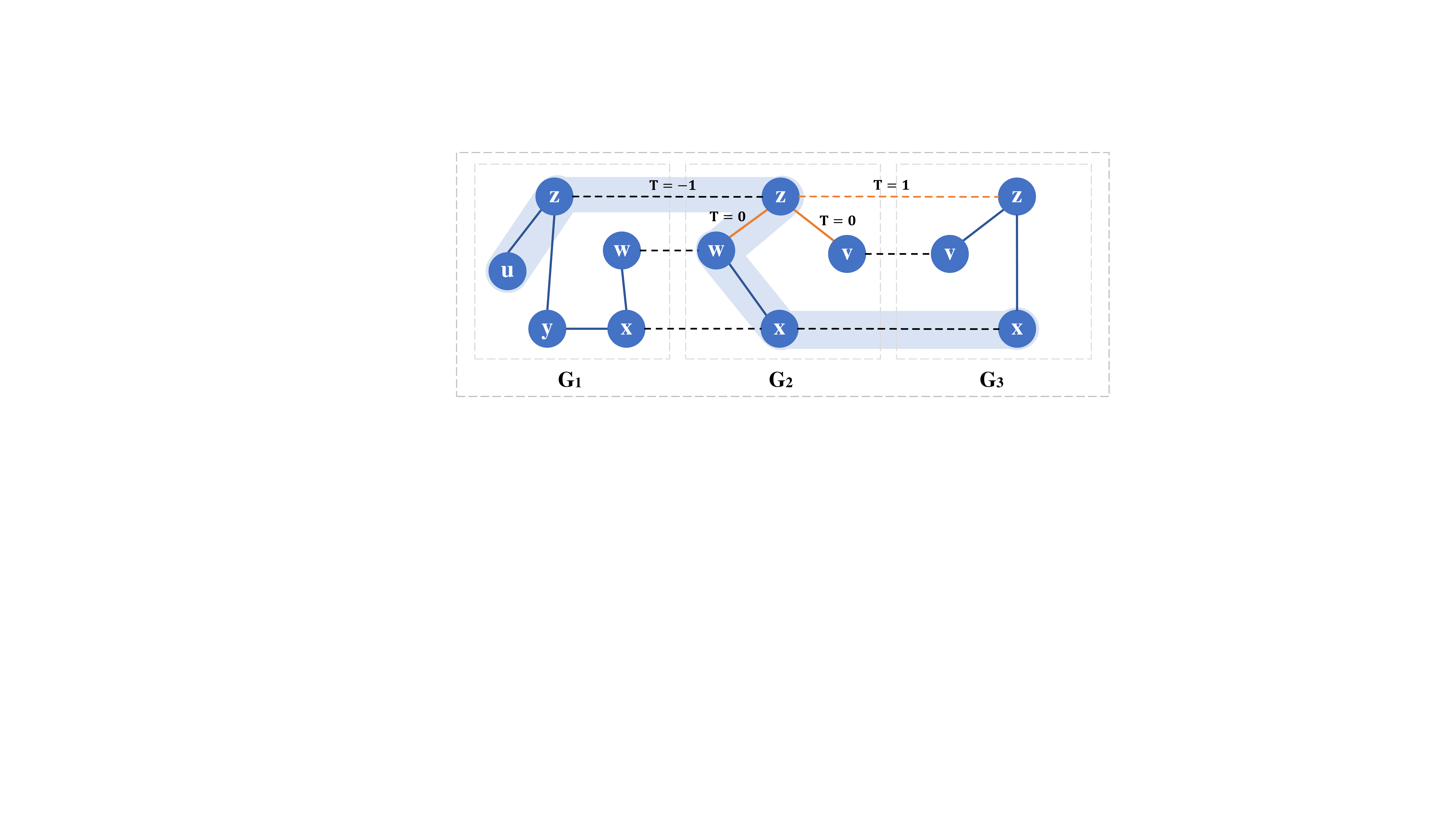}\caption{The blue block represents a valid temporal walk path starting in vertex $u$ in $G_{1}$. Accessible edges of a vertex $z$ in snapshot $G_{2}$ denoted as $L_t(z)$. Note that $L_t(z)$ are orange lines where accessible edges only appear in the current and nearby following snapshots.}
  \label{fig:sucessiveedges}
\end{figure}

% Now, the computational complexity in building TSSN is analyzed. The number of nodes in each snapshot and the time order in original network are represented by $|t|$ and $|v|$, respectively. According to the definition of TSSN, the time complexity in transforming the original network to TSSN is $\mathcal{O}(|t||v|)$. In general, constructing TSSN is time consuming. However, transforming the original network as TSSN has certain advantages, which will be explained in the following.
Self-connections in TSSN can make the random walk across successive snapshots, capturing the correlation between different snapshots, which may result in more appropriate embedding. For simplicity, we define a four-tuple $e = (u, v, w, t)$ in TSSN: for $\forall e \in E$, $Src(e) = u$, $Dst(e) = v$, $W(e) = w$, $T(e) = t$, where $u$ is the source vertex, $v$ is the target vertex, $w$ is the weight~(number of emails) and $t$ is the time accessibility. Let $\eta_{+} : \mathbb{R} \to \mathbb{Z}^{+}$ be a function that maps each vertex to an index based on the time order $t$, i.e., for a given vertex $u$ in snapshot $G_{i}$, we have $\eta_{+}(u) = i$. Therefore, we can design a sign function for each link in TSSN: $T(e) = \eta_{+}(v) - \eta_{+}(u) \in \{-1, 0, 1\}$, where $v$ is the target vertex and $u$ is the source vertex, to define the time accessibility of $v$ from $u$, i.e., $v$ is time accessible from $u$ if and only if the corresponding $T(e)\geq{0}$. Now we can define the \emph{temporal walk} as follows. 

%In each snapshot of TSSN, edges occur over a time span $\epsilon \subseteq \mathbb{T}$ where $\mathbb{T}$ is the temporal domain. When networks evolves, modeling temporal networks as TSSN fashion makes it completely simple to capture the edges and vertices' change (add or remove). In the proposed network, a valid walk is defined as a sequence of vertices connected by edges with non-decreasing time information $T(e)$. Here, we define the concept of a valid temporal walk as follows:
\begin{myDef}[\textbf{Temporal Walk}]
In TSSN, a temporal walk from vertex $v_{1}$ to vertex $v_{l}$ is an $l$-length sequence of vertices together with a sequence of $(l-1)$ edges $\{e_{1}, e_{2}, \cdots, e_{l-1}\}$, where $Src(e_{i}) = v_{i}$, $Dst(e_{i}) = v_{i+1}$, and $T(e_{i}) \geq 0$, for $1 \leq i \leq (l-1)$.
\end{myDef}

%A temporal walk represents a temporally valid sequence of edges traversed in non-decreasing order of edge's time information and therefore respect time. The definition of valid temporal walk in TSSN echoes the standard definition of the walk in static networks but with an additional constraint that edges traverse must obey time. See Fig. \ref{fig:sucessiveedges} for intuition. Any method that uses temporally invalid walks or static snapshot graphs without correlation is said to have temporal loss. 
In order to sample such temporal walks, we further define \emph{accessible edge} in TSSN as follows. 

\begin{myDef}[\textbf{Accessible Edge}]
  Given a TSSN ${G=(V,E)}$, the set of accessible edges for a vertex $v$ is defined as:
  \begin{equation*}\label{eq:successiveedges}
  L_{t}(v) = \{e\,|\,Src(e)=v, T(e) \geq 0\}
  \end{equation*}
  \end{myDef}
An example of accessible edge and temporal walk is presented in Fig.~\ref{fig:sucessiveedges}. Next, we will introduce different sampling biases by formulating the selection probability for each accessible edge $e \in L_{t}(v)$, and present a sampling strategy by combining these biases.  
%For each accessible edge $e \in L_{t}(v)$, we have temporal neighbor $Dst(e)$ of a vertex $v$. Observe that the $v$'s self-vertices appear multiple times in $L_{t}(v)$ since temporal edges can exist between the same vertice in two consecutive snapshots. This is another advantage to TSSN compared to previous work that connected edges from self-vertices make correlations between different snapshots. See Fig. \ref{fig:sucessiveedges} for an example. The set of successive edges plays the role of candidate for walkers to select possible successors. Therefore, we describe different sampling biases by formulating the selection probability for each successive edge $e \in L_{t}(v)$ and present a sampling strategy combining these selection biases. 

\subsection{Temporal Biased Walk}\label{sec:temporalwalk}
Based on the definitions in Sec.~\ref{sec:definition}, we design a second-order neighborhood sampling strategy $s$ to choose accessible edges. The search strategy is the joint transition probability we proposed, which is composed of the static edge weight, the structural transition probability, the temporal transition probability, and the role-based transition probability. %between the source vertice and it's first-order temporal neighbors.

Traditionally, for a given vertex $v$, we can perform a simple random walk. Let $Dst(e_{i})$ denotes the $i$th vertex in the temporal walk sequence $N_{s}(v)$ and random walk resides at vertex $c$. Thus each accessible edge $e \in L_{t}(c)$ can be assigned the selection probability:
\begin{equation}
P(e) = \frac{W(e)}{\sum_{e^{'} \in L_{t}(c)}W(e^{'})}
\end{equation}
where $W(e)$ is the weight between vertex $c$ and its temporal neighbor $x$, and $L_{t}(c)$ denotes the set of accessible edges of for vertex $c$. As illustrated in Sec.~\ref{sec:ossdataset}, interactions~(edge weights) exhibit some differences per role in the Apache email network. Thus we can use this simplest way to bias our temporal random walk, which is to sample the next vertex based on the static edge weight $W(e)$. 

However, this simple way does not explain the network structure nor can it help us explore different types of neighbors in the whole network. When the links are relatively sparse in the network, this strategy may get affected easily. Therefore, we propose a second-order temporal random walk method. In the proposed method, we introduce a joint transfer probability, which is composed of the static edge weight, the structural transition probability, the temporal transition probability, and the role-based transition probability for each source vertex's valid accessible edges. Suppose there is a random walk resides at vertex $c$, and the last traversed vertex is $t$. We calculate the latter three transition probabilities of the vertex $c$'s temporal accessible edges as follows.
% The Node2vec \cite{grover2016node2vec} method develops a random walk that can explore neighbors in a mixed fashion of BFS and DFS. Nevertheless, this strategy ignores important temporal information, which contains the evolution of the network structure, and thus may lead to incomplete network representations.
% For a given vertice $v$ we can divide its first-order neighbors into three categories according to vertice's structure in the network: inward vertices, outward vertices, and return vertice. The return vertice denotes the last traversed vertice. The definition of an inward (or outward) vertice is based on whether there is a link with the return vertice. If the vertice has a link with the return vertice, it is an inward vertice, otherwise an outward vertice. Different types of vertices play an important role in the network, thus we need to take vertices' roles structurally into consideration for each vertice's next walk. In addition, when performing walks, the correlation in different snapshots is considerable important, it can reflect the evolution of the network structure. Further, we assume that roles' real identities have a positive impact on search strategy, i.e., we can explore that whether the search direction of source vertice is more inclined to the neighbors who have the same or different label. The search strategy combined with these assumptions may result in more informative embeddings.

\textbf{Structural Transition Probability:} We define the structural transition probability with return parameter $r$ and in-out parameter $q$ similar to~\cite{grover2016node2vec}.
%  Considering a random walk now is stopping at vertice $c$, $t$ is the last traversed vertex, and the edge between vertices $c$ and $x$ is the valid accessible edge of vertex $c$. 
For each valid accessible edge $e \in L_{t}(c)$, we set the unnormalized structural transition probability to $P_{S}(e) = \psi_{S}(e) \cdot W(e)$ with
\begin{equation}
\psi_{S}(e) =  \left\{
\begin{array}{rcl}
  1/q,      &     &d_{tx}=2\\
  1,        &     &d_{tx}=1\\
  1/r,      &     &d_{tx}=0
\end{array} \right. 
\end{equation}
where $d_{tx} \in \{0,1,2\}$ denotes the shortest path distance between vertex $t$ and $x$. 

It is worth noticing that the initial in-out parameter $q$ and return parameter $r$ jointly determine each vertex's search direction. Our method, like~\cite{grover2016node2vec},~\cite{chen2019pso}, uses the return parameter $r$ and in-out parameter $q$ to control how fast the walk explores and leaves the neighborhood of starting vertex. %Thus our method can explore more different types of vertices to get more informative embeddings.

\textbf{Temporal Transition Probability:} Apart from the structural features, the temporal information also counts for much in the vertex representation learning. When we divide the whole network into different snapshots based on time span, each snapshot represents a part of the network structure and the gradual change of time slice reflects the evolution process of network. Ignoring the correlation information that exists between two snapshots at consecutive time steps may cause the loss of temporal information. Hence, we propose temporal transition probability to capture vertex's behavior changes in different snapshots. In this case, the probability of selecting each edge $e \in L_{t}(c)$ can be given as:
\begin{equation}
  P_{T}(e) = \frac{\psi_{T}(e)}{\sum_{e^{'} \in L_{t}(c)}\psi_{T}(e^{'})},
\end{equation}
where $\psi_{T}(e)$ is expressed as
\begin{equation}
  % \alpha(e) = \frac{exp(T(e))}{\sum_{e^{'} \in L(c)}exp(T(e^{'}))}
  \psi_{T}(e) =  \left\{
\begin{array}{rcl} 
  \alpha,     &     &T(e) > 0\\
  1-\alpha,   &     &T(e) = 0
\end{array} \right. 
\end{equation}
Here, the temporal bias $\alpha$ $(0.1 \le \alpha \le 0.9)$ decides whether the temporal walk resides on the current snapshot or transfers to the next. 

\textbf{Role-Based Transition Probability:} Every role has a different communication tendency as illustrated in Sec.~\ref{sec:preliminaryanalysis}, which means that the individuals may be more inclined to communicate with those of the same role or the opposite. To explore more various temporal neighbors of a role via communication tendency, we consider both unbiased and biased sampling strategies as follows.
\begin{itemize}
  \item \emph{Role Unbiased Sampling (RUS)}. It assumes that each accessible edge $e \in L_{t}(c)$ of vertex $c$ has the same probability to be sampled: 
  % This is the default setting in this paper. 
  \begin{equation}
    P_{R}(e) = \frac{1}{|L_{t}(c)|}
  \end{equation}
  \item \emph{Role Biased Sampling (RBS)}. We have role bias parameter $\beta$ $(0.1 \le \beta \le 0.9)$ to control whether the temporal walk is toward the same type of vertices or different. The biased transition probability of each $e \in L_{t}(c)$ is then defined as: 
  \begin{equation}
    P_{R}(e) = \frac{\psi_{R}(e)}{\sum_{e^{'} \in L_{t}(c)}\psi_{R}(e^{'})},
  \end{equation}
where $\psi_{R}(e)$ is set to 
\begin{equation} 
  \psi_{R}(e) =  \left\{
\begin{array}{rcl} 
  \beta,     &     &\omega(t) = \omega(x)\\
  1-\beta,   &     &\omega(t) \neq \omega(x)
\end{array} \right. 
\end{equation}
with $\omega(v)$ denoting vertex $v$'s real identity, i.e., user or developer, $t$ being the last traversed vertex, and $x=Dst(e)$. 
\end{itemize}

\textbf{Joint Transition Probability:} Now, we normalize the aforementioned structural transition probability, temporal transition probability, and role-based transition probability, and then combine them as one. Finally, each edge $e \in L_{t}(c)$ can be assigned the selection probability:
\begin{equation}
  P(e) = P_{S}(e)\,P_{T}(e)\,P_{R}\,(e).
\end{equation}

Based on the joint transition probability, we propose a second-order neighborhood sampling strategy $s$ which can help each vertex find a suitable search direction and get its optimal temporal accessible edges. In each vertex's temporal walk, the in-out parameter $q$, return parameter $r$, temporal bias $\alpha$ and role bias $\beta$ jointly determine the search direction. 

The return parameter $r$ mainly controls the probability of the source vertex revisiting the last traversed vertex. When $r$ is small, it would keep the walk close to the source vertex. On the other hand, setting it to a large value ensures that the walk is less likely to be the already visited vertices. The parameter $q$ prefers to consider searching for different types of inward and outward vertices structurally. The definition of an inward~(or outward) vertice is based on whether there is a link with the last traversed vertice. When $q>1$, the next walk of the source vertex is more inclined to return to the source vertex, which is more like a local exploration like the BFS behavior. When $q<1$, the next vertex is more likely to walk away from the source vertex. This method can make the source vertex explore a wider range of vertices, which is a kind of approximate DFS behavior. By adjusting the parameter $q$, we allow our search direction to combine BFS with DFS. On the whole, the in-out parameter $q$ and return parameter $r$ control the search direction in spatial domain simultaneously. 

Temporal bias $\alpha$ decides the temporal search orientation: resides on current snapshot or move to next snapshot. If $\alpha$ is small, the temporal walk is more inclined to stay in the current snapshot, otherwise the walk favors edges appearing in future snapshot. This helps to explore changes in vertices' interaction during very different time periods as the network evolves. Role bias $\beta$ control vertex's communication tendency. If $\beta$ is large, the temporal walk is more likely to traverse the same type of vertices as the source vertex, otherwise the walk encourages the exploration of vertices of different type. %These samplings are more helpful to search for various types of roles both in the spatial and temporal domains. 

\subsection{Learning Temporal Network Embeddings}
Our goal is to obtain a mapping function $f:V \to \mathbb {R}^{d}$, which maps a given vertex to a $d$-dimensional representation. For a vertex $v \in V$, let $N_{s}(v)$ denotes the set of temporal neighbors that are generated according to the search strategy $s$, and $f_{t}(v)$ is the representation of vertex $v$ in snapshot $G_{t}$. Our objective function maximizes the log-probability of observing $N_{s}(v)$ and historical embedding $f_{t}(v)$ for the vertex $v$ conditioned on its representation:
\begin{equation}\label{eq:1}
\max_{f}\sum_{v \in V}\log(Pr(N_{s}(v),f_{t}(v)\,|\,f(v))).
\end{equation}
We assume that the temporal neighbors in $N_{s}(v)$ and the vertex's historical representations $f_{t}(v)$ are independent of each other. Accordingly, we factorize the formula: 
\begin{equation}
\label{eq:2}
\begin{aligned}
&\log(Pr(N_{s}(v),f_{t}(v)\,|\,f(v))) \\
=&\log(\prod_{u_{i} \in N_{s}(v)}Pr(u_{i}\,|\,f(v))) + \log(Pr(f_{t}(v)\,|\,f(v))).
\end{aligned}
\end{equation}

Based on the network analysis, we can see that the likelihood of observing a source vertex is independent of observing any other and the definition of neighborhood vertices is symmetric~\cite{grover2016node2vec}. Therefore, we factorize the likelihood of observing temporal neighbors and model the likelihood of every source-neighborhood vertex pair as a softmax unit that is parametrized by a dot product of their mapping features. Learning representations using random walk has proved to measure better graph proximity, and thereby improving the performance~\cite{hamilton2017representation},~\cite{goyal2018graph}. Hence, we use random walk to learn the conditional probability of observing a vertex $u_{i}$ given the learned representation $f(v)$ as follows:
\begin{equation}
\label{eq:3}
Pr(u_{i}\,|\,f(v))=\frac{\exp(f(u_{i})\,f(v))}{\sum_{n \in V}\exp(f(n)\,f(v))},
\end{equation}
where $u_{i} \in N_{s}(v)$ is the $i$th neighbor of vertex $v$. With the above hypothesis, the objective function in Eq.~(\ref{eq:1}) can be described as:
\begin{equation}
\label{eq:4}
\begin{aligned}
\max_{f}\sum_{v \in V}\log(\prod_{u_{i} \in N_{s}(v)} &\frac{\exp(f(u_{i})\,f(v))}{\sum_{n \in V}\exp(f(n)\,f(v))}) \\
&+\log(Pr(f_{t}(v)\,|\,f(v))).
\end{aligned}
\end{equation}
Considering the complexity of this objective function, we use negative sampling strategy to approximate it~\cite{mikolov2013distributed}. The stochastic gradient descent~(SGD)~\cite{bottou2010large} method is used to iteratively update the objective function. 

Due to the nonlinear nature of real-world networks, we define a novel search strategy $s$ that samples different temporal neighbors of a given source vertex $v$. The temporal neighbors $N_{s}(v)$ are not restricted to just nearest neighbors but also have vastly structural similarity with the source vertex in spatial and temporal domains simultaneously. While the above seems to just consider the process of network topological properties, it actually takes into account the role's real identity to get more informative representations.

% it actually takes into account the structure of network at different times and thus reflects the evolution process of network topology properties. 

\begin{algorithm}[htbp] 
  \caption{\textbf{Partner Recommendation}}
  \label{alg:1}
  \KwIn{temporal graph $G=(V,E)$, return $r$, in-out $q$, \\
  temporal bias $\alpha$, role bias $\beta$, time span $\epsilon$, dimension $d$,
  walks per vertex $w$, walk length $l$, window size $k$}
  \KwOut {$f(v)$ for $\forall v \in V$}
  Initialize set of temporal walks $N_{s}$ to $\emptyset$\\
  $G^{'}$ = CreateTSSN($G$, $\epsilon$)\\
  
  \For {iter=1 to w}
  {
    \For {all vertex $v \in V$}
    {
      $P$ = PrecomputeTransitionProbability($G^{'}$, $r$, $q$, $\alpha$, $\beta$)\\
      $walk$ = TemporalBiasedWalk($G^{'}$, $v$, $l$, $P$) \\
      Append $walk$ to $N_{s}$
    }
  }
  $f$ = StochasticGradientDescent($k$, $d$, $N_{s}$)\\
  return $f \in \mathbb{R}^{|V| \times d} $
\end{algorithm}

\begin{algorithm}[htbp] 
\caption{\textbf{Temporal Biased Walk}}
\label{alg:2}
\KwIn{time-series snapshot network $G^{'}$, start vertex $u$, walk length $l$, transition probability $P$}
\KwOut {temporal walk $walk$}
Initialize $walk$ to [$u$], $walk_{e}$ to $\emptyset$\\
\For {iter=1 to $l$}
{ 
  \If{$len(walk) == 1$}{  
    $curr = walk[-1]$ \\
    $e$ = AliasNodeSample($curr$, $P$) \\
    Append $e$ to $walk_{e}$, append $Dst(e)$ to $walk$\\}
  \Else{
    $curr_{e} = walk_{e}[-1]$ \\
    $e$ = AliasEdgeSample($curr_{e}$, $P$) \\
    Append $e$ to $walk_{e}$, append $Dst(e)$ to $walk$\\
  }
}
% Remove $walk$ adjacent duplicate vertices \\
return $walk$
\end{algorithm}

In Algorithm~\ref{alg:1}, we propose the framework to learn time-preserving embeddings in TSSN. Our procedure in Algorithm~\ref{alg:1} generalizes the Skip-Gram architecture to learn time-series snapshot network embeddings. In this biased random walk, every start vertex has a unique search strategy. The three phases of temporal biased walk~(Algorithm~\ref{alg:2}), i.e., preprocessing to compute joint transition probability, random walk simulations, and optimization using SGD, are executed sequentially. Each phase is parallelizable and can be executed asynchronously, which contributes to the overall scalability of TBW. Furthermore, TBW can be easily used for other deep graph models since the temporal walks can serve as input vectors for neural networks. There are many random walk methods that can be adapted in TSSN because it is not tied to any particular approach.  

% Now, we analyze the corresponding computational complexity. The number of nodes in each snapshot and the time order in the original network are represented by $|V|$ and $t$, respectively. According to the definition of TSSN, the time complexity in transforming the original network to TSSN is at most $\mathcal{O}(t \cdot |V|)$. For each vertex visited in a random walk of maximum length $l$, Algorithm~\ref{alg:2} needs to consider all neighbors of that vertex and then update the transition probabilities. Thus, the time for both accessing and updating transition probabilities is in $\mathcal{O}(|V| \cdot l)$. Therefore, the computational complexity for vertices in all snapshots is around $\mathcal{O}( l \cdot |V| \cdot t)$.
Now, we analyze the corresponding computational complexity. The average  number of vertices in each snapshot and the time order in the original network are represented by $|\overline{V}|$ and $t$, respectively. According to the definition of TSSN, the time complexity in transforming the original network to TSSN is at most $\mathcal{O}(t \cdot |\overline{V}|)$. Let $\overline{D}$ be the average degree of each snapshot. For each vertex visited in a random walk of maximum length $l$, Algorithm~\ref{alg:1} needs to consider all neighbors of that vertex and then update transition probabilities. Therefore, the time for both accessing and updating transition probabilities is $\mathcal{O}(l \cdot (\overline{D} + |\overline{V}|))$ and  the time complexity for vertices in all snapshots is around $\mathcal{O}(l \cdot (\overline{D} + |\overline{V}|)) \cdot t) $.

\section{Experiments on Apache}\label{sec:experiments}
\subsection{Experiment Setup}
We compare the performance of TBW with six random walk based network embedding methods. And the basic settings are described as follows:

\begin{itemize}
  \item \textbf{LINE}~\cite{tang2015line} preserves both the local and global network structures through modeling vertex co-occurrence probability and conditional probability. The final representation for each vertex in this work is created by second-order representation.
  \item \textbf{BiasedWalk}~\cite{nguyen2018biasedwalk} is a random walk based sampling method, which can behave as Breath-First-Search~(BFS) and Depth-First-Search~(DFS) sampling, in order to capture the homogeneity and role equivalence between vertices in the network. We set the parameters to the provided defaults.
  \item \textbf{DeepWalk}~\cite{perozzi2014deepwalk} is the first vertex embedding method that obtains vertices context via random walks, which uses Skip-Gram model and uniform random walks to learn the neighborhood structure of the graph. 
  \item \textbf{Node2vec}~\cite{grover2016node2vec} keeps the neighborhood of vertices to learn the vertex representation in the network, and it achieves a balance between homophily and structural equivalence. The ranges of its hyper parameters in this paper are set to $p,q\in\{0.5, 1, 2\}$.
  \item \textbf{Dynnode2vec}~\cite{mahdavi2018dynnode2vec} is an embedding method using temporal information based on~\cite{grover2016node2vec} that can capture evolving patterns in temporal networks. It uses evolving random walks and initializes the current graph embedding with previous embedding vectors.
  \item \textbf{CTDNE}~\cite{nguyen2018continuous} is a general framework for incorporating temporal information into network embedding methods, which is based on random walk and stipulates that the timestamp of the next edge in the walk must be larger than that of the current edge. 
\end{itemize}

We take one month as the time span to construct TSSN for each dataset and conduct our partner recommendation experiments with TBW. For all random walk based embedding methods, including ours, we utilize the same hyper-parameter setting~(the number of walks per vertex $w = 10$, the length of walk $l = 80$, and the size of context window $k = 5$). To guarantee $d \ll D$~($D$ is the number of network vertices), we set the dimension $d = 128$ for all datasets and methods. After the embeddings are learned for each vertex, we use average operation on the learned embedding vectors of pairwise vertices to compute the feature vector for the corresponding edge. For all baselines, we implement experiments by using a one-vs-rest logistic regression classifier with hold-out validation of 25\% on Apache datasets. Experiments are repeated for 10 random seed initializations and the average performance~(AUC) is reported. In the following experiments, our hyper-parameters set as follows: return parameter $r$ and in-out parameter $q$ are grid searched in $\{0.5, 1, 2\}$, temporal bias $\alpha$ and role bias $\beta$ vary in $[0.1, 0.9]$.

\subsection{Partner Recommendation}
% In this subsection, we discuss the question which represents the key application of our study: How accurately TBW makes role recommendation from Apache email data?
\begin{table*}[tbp]
  \renewcommand{\arraystretch}{1.2}
  \setlength{\abovecaptionskip}{0pt}
  \setlength{\belowcaptionskip}{5pt}
  \centering
  \caption{AUC scores for traditional partner recommendation across different roles.}
    \begin{tabular}{c|ccccccc}
    \midrule
    \midrule
           Project    & LINE  & BiasedWalk  & DeepWalk & Node2vec & Dynnode2vec & CTDNE & TBW \\
    \midrule
           Airflow    & 0.8214  & 0.8645    & 0.8631   & 0.8909   & 0.8671      & 0.8540 & \textbf{0.9211 } \\
           Beam       & 0.9072  & 0.9243    & 0.9146   & 0.9197   & 0.9102      & 0.9420 & \textbf{0.9526 } \\
           CarbonData & 0.8631  & 0.8613    & 0.8387   & 0.8669   & 0.8546      & 0.7668 &\textbf{0.9060  } \\
           Cordova    & 0.8765  & 0.9308    & 0.9345   & 0.9371   & 0.9287      & 0.9267 & \textbf{0.9536 } \\
           Geode      & 0.8316  & 0.8598    & 0.8551   & 0.8468   & 0.8657      & 0.8776 & \textbf{0.8947 } \\
           HAWQ       & 0.8175  & 0.8125    & 0.7943   & 0.8029   & 0.8511      & 0.8681 & \textbf{0.8842 } \\
           Impala     & 0.8512  & 0.8672    & 0.8479   & 0.8643   & 0.8626      & 0.8564 & \textbf{0.9041 } \\
           Metron     & 0.8393  & 0.8072    & 0.8158   & 0.7828   & 0.8466      & 0.7849 & \textbf{0.8740 } \\
           Spark      & 0.8000  & 0.8198    & 0.8167   & 0.8362   & 0.8305      & 0.8277 & \textbf{0.8670 } \\
           Zeppelin   & 0.8379  & 0.8996    & 0.8935   & 0.8945   & 0.9042      & 0.9025 & \textbf{0.9365 } \\ \midrule
           Avg        & 0.8446 	& 0.8647 	  & 0.8574 	 & 0.8642 	& 0.8721      & 0.8607 & \textbf{0.9094 } \\
    \midrule
    \midrule
    \end{tabular}
  \label{tab:rolerecommendation}
\end{table*}

\begin{table*}[!t]
  \renewcommand{\arraystretch}{1.2}
  \setlength{\abovecaptionskip}{0pt}
  \setlength{\belowcaptionskip}{5pt}
  \centering
  \caption{AUC scores for time-preserving partner recommendation across different roles.}
    \begin{tabular}{c|ccccccc}
    \midrule
    \midrule
    Project    & LINE   & BiasedWalk & DeepWalk & Node2vec & Dynnode2vec  & CTDNE  & TBW \\
    \midrule
    Airflow    & 0.8546 & 0.8882     & 0.9000   & 0.9189   & 0.9309       & 0.8513 & \textbf{0.9404} \\
    Beam       & 0.9319 & 0.8880     & 0.8497   & 0.8885   & 0.9420       & 0.8542 & \textbf{0.9513} \\
    CarbonData & 0.8354 & 0.8244     & 0.8085   & 0.8462   & 0.8751       & 0.8665 & \textbf{0.9040} \\
    Cordova    & 0.8952 & 0.9383     & 0.9423   & 0.9548   & 0.9632       & \textbf{0.9679} & 0.9577 \\
    Geode      & 0.8922 & 0.9125     & 0.8949   & 0.9119   & 0.9237       & 0.9378 & \textbf{0.9396} \\
    HAWQ       & 0.7616 & 0.7333     & 0.6945   & 0.7634   & 0.8071       & 0.8441 & \textbf{0.8607} \\
    Impala     & 0.8621 & 0.7506     & 0.5945   & 0.7267   & 0.8507       & 0.7885 & \textbf{0.8838} \\
    Metron     & 0.8947 & 0.8736     & 0.8367   & 0.8699   & 0.9265       & 0.8040 & \textbf{0.9400} \\
    Spark      & 0.7771 & 0.8415     & 0.8191   & 0.8449   & 0.8609       & 0.7248 & \textbf{0.8838} \\
    Zeppelin   & 0.8679 & 0.9152     & 0.8898   & 0.9138   & 0.9352       & \textbf{0.9551} & 0.9379  \\ 
    \midrule
    Avg        & 0.8573 & 0.8566 	   & 0.8230 	& 0.8639 	 & 0.9015       &	0.8594 	& \textbf{0.9199}     \\
    \midrule
    \midrule
    \end{tabular}
  \label{tab:time-preservingrolerecommendation}
\end{table*}

\begin{figure*}[!t]
  \centering
  \includegraphics[width=\linewidth]{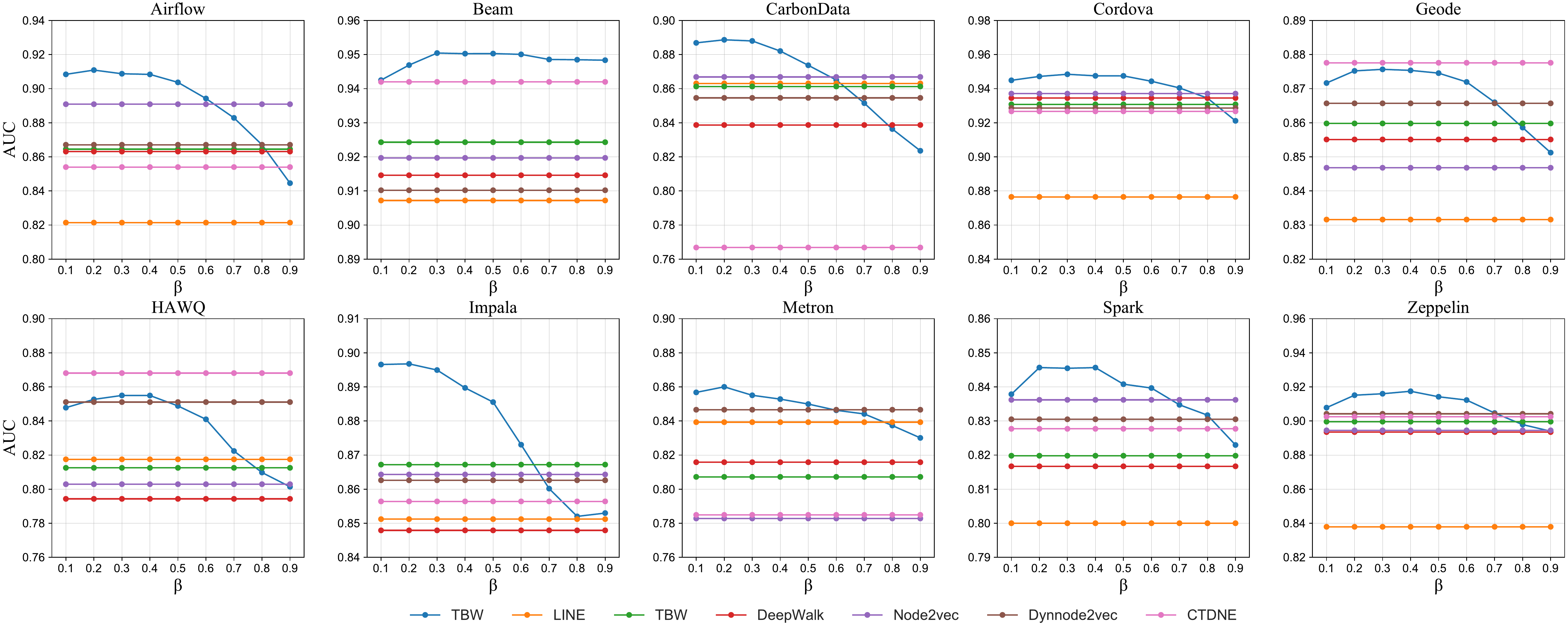}
  \caption{The performance of TBW as function of role bias $\beta$ which varies in $[0.1, 0.9]$.}
  \label{fig:rolescommunicationtendency}
\end{figure*} 

\begin{figure*}[!t]
  \centering
  \includegraphics[width=\linewidth]{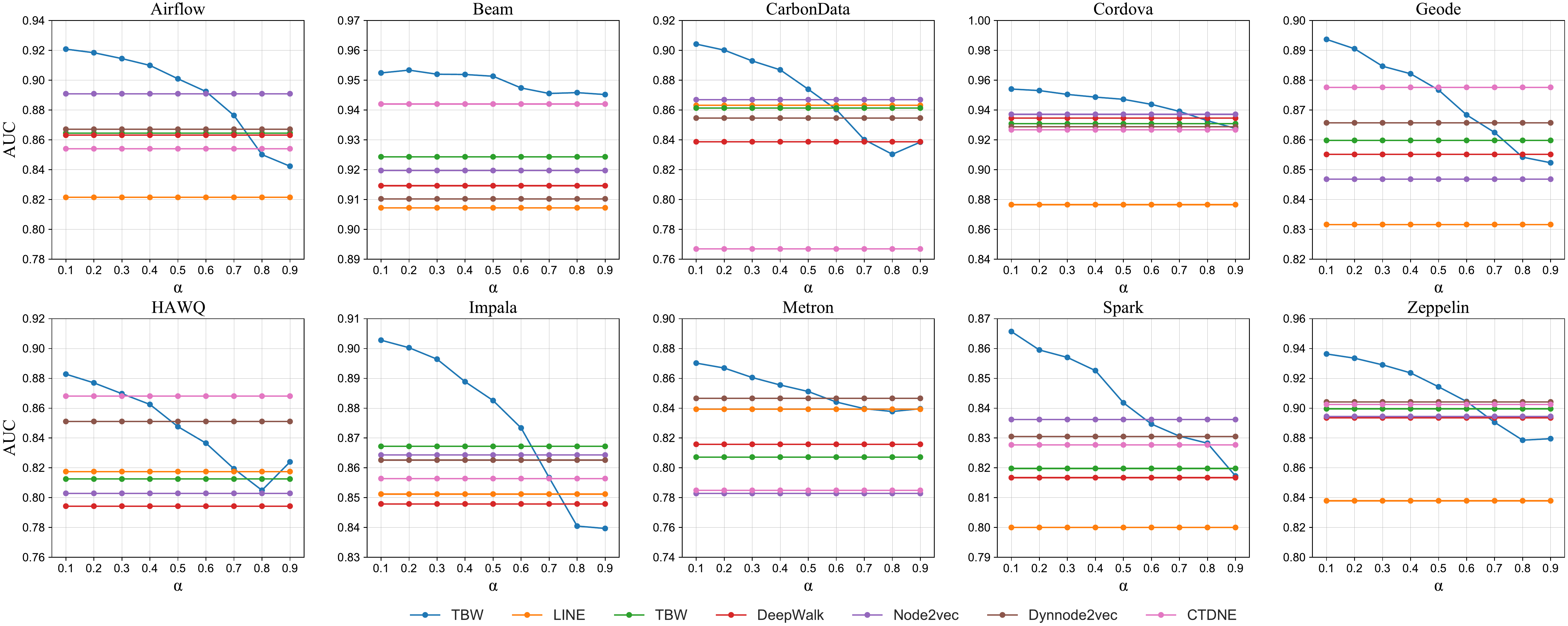}
  \caption{The performance of TBW as function of temporal bias $\alpha$ which varies in $[0.1, 0.9]$.}
  \label{fig:alpha}
\end{figure*} 

We treat the partner recommendation problem as a link prediction task between two roles in the Apache community: users and developers. We adopt the following recommendation strategy to investigate the desirability for those in need, i.e., for a given user/developer, we make a random recommendation to the target without considering their roles. Partner recommendation can be divided into traditional partner recommendation and time-preserving partner recommendation, i.e., traditional link prediction and temporal link prediction. The traditional partner recommendation refers to the use of all information to recommend partners, while time-preserving partner recommendation refers to the use of historical information for current participants to make recommendations, which is more practical. Before the experiments, we hide a certain fraction of individuals' connections in the email network for a given project, and our goal is to predict these missing connections so as to achieve the recommendation via link prediction.

For traditional partner recommendation, we first randomly hide 25\% of links in the original network as the ground truth and use the remaining to train all network embedding models. The test set consists of the hidden 25\% links in the original network as positive samples, and the same number of disconnected vertex pairs are randomly selected as negative samples. Table~\ref{tab:rolerecommendation} shows the performance of all the compared methods on traditional partner recommendation. To make the model more practical, we use the historical data to predict future edges~(time-preserving partner recommendation), i.e., we first sort the edges by time~(ascending) and use the first 75\% to create email network for the representation learning. The remaining 25\% are considered as positive samples and an equal number of disconnected vertex pairs are randomly chosen as negative samples. The time-preserving partner recommendation results are given in Table~\ref{tab:time-preservingrolerecommendation}.

From Table~\ref{tab:rolerecommendation} and Table~\ref{tab:time-preservingrolerecommendation}, we can observe that, generally, temporal methods achieve significantly better performance over static baselines in most cases, which is reasonable since static methods totally ignore temporal information. However, baseline temporal methods may also perform poorly in some projects with non-uniform distribution of activities while our method is relatively stable across different projects. In fact, our TBW performs the best in most cases with the average AUC above 0.9 for both traditional and time-preserving partner recommendation. Such superiority is partly due to the role information integrated in our method, which help TBW learn appropriate and meaningful user/developer behavioral characteristics. Moreover, the outstanding performance of TBW also suggests that our TSSN could be more informative than the traditional temporal network, which better combines spatial and temporal properties and thus improves TBW to certain extent. It is worth noting that our TSSN model is quite general, i.e., many other random walk based approaches can also be generalized by using our proposed TSSN~\cite{cavallari2017learning},~\cite{dong2017metapath2vec},~\cite{ribeiro2017struc2vec}, and can be applied in many other applications beyond OSS.

\subsection{Parameter Sensitivity}  
Here, we mainly focus on the effects of role bias $\beta$ and temporal bias $\alpha$ on the performance of TBW, since these two parameters are newly introduced in our method. The other parameters are set as follows: the in–out parameter $q$ and the return parameter $r$ are grid searched in $\{0.5, 1, 2\}$, the number of walks $w = 10$, the walk length $l = 80$, $\alpha = 0.5$, $\beta = 0$. 
% The experiment was repeated 30 times to avoid errors and the average performance~(AUC) is reported.

As shown in Fig.~\ref{fig:labeltrend}, we find that individuals in most projects are more likely to contact those of different roles. Thus we first investigate the effect of role bias $\beta$ which varies in $[0.1, 0.9]$. The results of random recommendation are shown in Fig.~\ref{fig:rolescommunicationtendency}, where we can see that, generally, the link prediction performance steadily improves as role bias $\beta$ decreases. Since smaller $\beta$ encourages exploration of multiple types of vertices, this result validates again that, in OSS projects, users always seek developers for help, while developers are inclined to share knowledge with users. Therefore, we need choose smaller role bias $\beta$ to address this tendency, so as to improve the recommendation performance. 

Besides, we further vary the temporal bias $\alpha$ in $[0.1, 0.9]$, to investigate how this parameter influences the recommendation performance. The results are shown in Fig.~\ref{fig:alpha}, where we can see that the link prediction performance decreases as the temporal bias $\alpha$ increases. This may be because a larger $\alpha$ usually causes a faster transfer to the next snapshot, resulting in incomplete sampling in the current snapshot. Therefore, in this work, a small temporal bias is preferred to capture enough structural information in different snapshots, so as to ensure an acceptable recommendation result. 

More interestingly, we find that, for the projects CarbonData, HAWQ, and Metron, AUC increases as $\alpha$ changes from 0.8 to 0.9, reversing the general trend. We thus double check the data and find that these projects are relatively active, but there is little email activity in several months. Such non-uniform distribution of activities may lead to this abnormal phenomenon. An alternative way is to choose a fixed number of activities, rather than a fixed time window, to establish TSSN, so that the sizes of snapshot networks could be more similar, leading to a clearer trend. In fact, we adopt this method on the projects HAWQ, CarbonData, and Metron, and the above abnormal phenomenon indeed disappears.

%\begin{figure}[!t]
%  \centerings
%  \includegraphics[width=\linewidth]{image/alphatest.pdf}
%  \caption{The performance of TBW with establishing TSSN by a certain number of activities rather than a fixed time window.}
%  \label{fig:alphatest}
%\end{figure} 
%it is necessary to find a proper temporal bias to balance structural and temporal domains. In summary, all of these parameters have a relatively high impact on the performance of TBW, especially role bias $\beta$ and temporal bias $\alpha$  described in Sec.~\ref{sec:temporalwalk}.

% Generally speaking, we observe that the performance of TBW improves as the structural parameters $r$ and $q$ decrease in most cases. While a low $r$ encourages inward exploration, i.e., the walk does not go too far from the start vertex, a low $q$ ensures a wider exploration. The results show that temporal bias $\alpha$ increases have a negative impact on the performance of TBW. Usually, large $\alpha$ causes a fast transfer to the next snapshot, resulting in incomplete sampling in the current snapshot. Therefore, it is necessary to find a proper temporal bias to balance structural and temporal domains. We also examine how the number of walks $w$ and walk length $l$ affect the performance. We observe that performance tends to saturate once the number and length of walks per source improve in most cases. In summary, all of these parameters have a relatively high impact on the performance of TBW, especially those described in Sec.~\ref{sec:temporalwalk}.

\subsection{Insight after Partner Recommendation}
The above experimental results clearly show that our TSSN model together with TBW, which integrates temporal information, structural information, and role information, can significantly improve the performance of partner recommendation in OSS. Such precise partner recommendation is critical for the efficiency and flourishing of OSS community. On one hand, for those projects with more email activities, recommending partners to those in need can prevent them from searching in the information explosion mailing list; while on the other hand, for those projects with fewer email activities, making partner recommendation can stimulate developers to participate in further development, and thus may dramatically increase their productivity on the projects.

\begin{figure}[!t]
  \centering
  \includegraphics[width=0.8\linewidth]{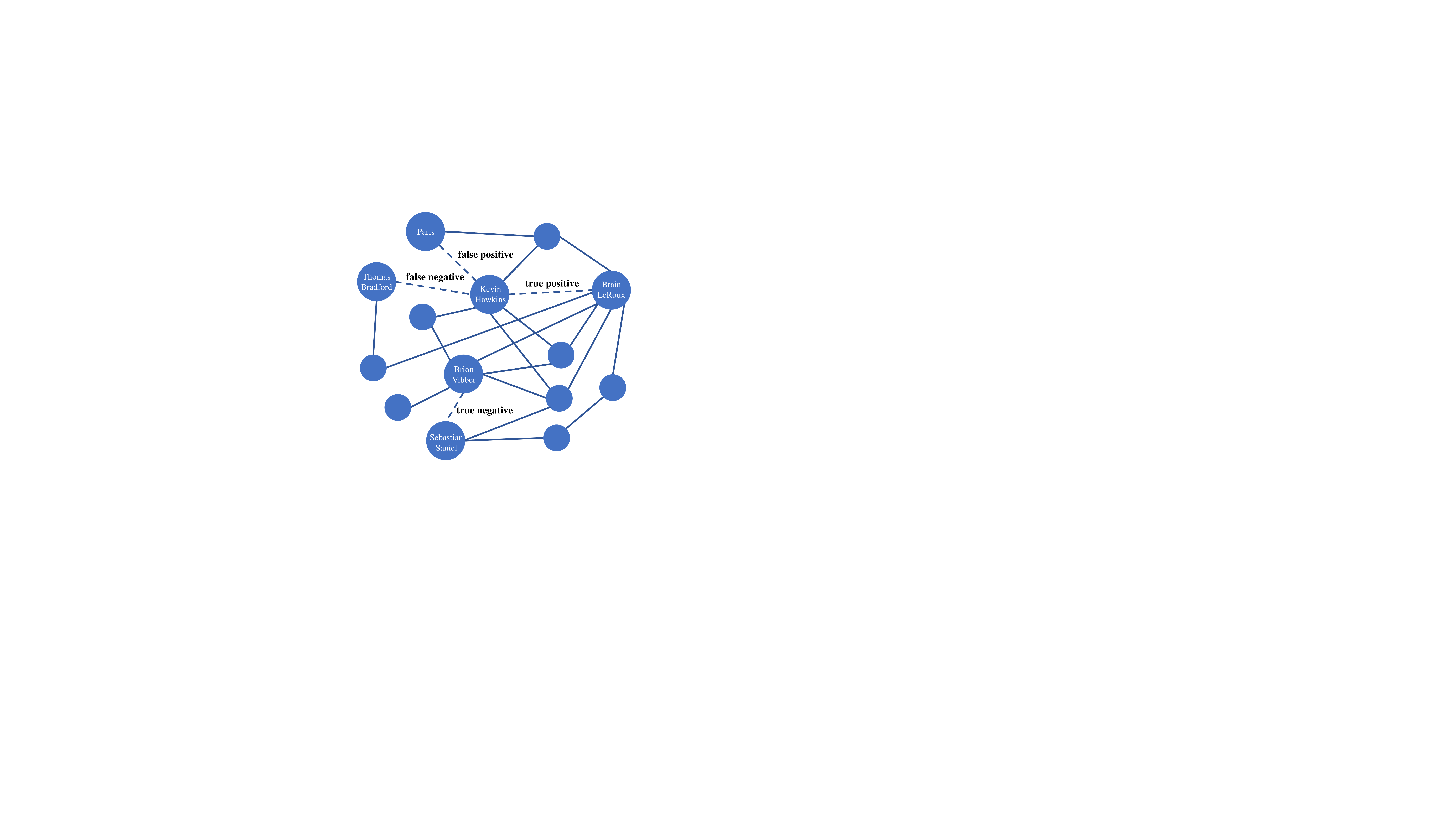}
  \caption{Some cases of behavioral similarity in project Cordova.}
  \label{fig:somecases}
\end{figure}

Now, we discuss some cases of behavioral similarity for true positive, false positive, true negative, and false negative, as shown in Fig.~\ref{fig:somecases}. For the case of true positive, we make successful partner recommendation for those individuals have similar behaviors. For instance, Kevin Hawkin and Brain LeRoux are both central individuals in the social network, i.e., several individuals ask them for help, and most of their communication targets are overlapping, thus we can recommend each other for them due to their structural similarity. For Paris and Kevin Hawkins, although there are email exchanges between them, our method cannot capture the similarity of their behaviors since Paris has too few email behaviors, resulting in false positive. For those disconnected individuals, e.g., Kevin Hawkins and Thomas Bradford, one is the central individual while the other is marginal. Their communication targets are quite different, resulting in the low behavioral similarity between them, and thus our method predicts there is no link between them. In fact, they indeed do not have email exchanges in reality, leading to false negative. For the case of true negative, Brion Vibber and Sebastian Saniel have one same communication target, since Sebastian Saniel only has two neighbors, our approach thinks Brion Vibber and Sebastian Saniel should be connected to each other, while in reality they are not.

\section{Threats to validity}~\label{sec:Threat}
First, only 10 OSS projects are selected from the same foundation. Although they contain most developers and email exchanges among all the graduated projects, such choice may limit the generalization of the results. Our proposed method therefore needs to be tested on a greater variety of OSS projects, and even other communities, to validate the generalization of TSSN model.

Second, we only consider email communication here, while in fact individuals may coordinate their work via other communication tools. Therefore, if possible, other types of communication could also be considered in future work to make the study more comprehensive.

Third, in OSS projects, participants may have different aliases and email addresses. To accurately scale an individual's activity, we introduce identity matching to merge different aliases. The proposed de-aliasing algorithm might introduce extra noises. Alternative de-aliasing methods~\cite{bird2006mining, wiese2016mailing} could be adopted together to further improve the matching precision so as to reduce noises.

However, such incompleteness of data, e.g., can't get email activity between all individuals or correctly identify an individual in multiple aliases, may influence all the embedding and link prediction methods. Therefore, the outstanding performance of our TSSN model in this paper still partly validates its general effectiveness, which could be further proved on more comprehensive datasets in the future.

% \textcolor{blue}{However, such incompleteness of data, e.g., can't get email activity between all individuals or correctly identify an individual in multiple aliases, may influence all the embedding and link prediction methods. Here discuss some cases of false positives and true negatives for the behavioral similarity, i.e., failed to recommend partners to individuals who should have email exchanges or two unrelated individuals. For example, in project Cordova, Paris Stamatopoulos and Kevin Hawkins email with the same group of individuals, and thus they have the same topics of interest and email exchanges. Since Kevin Hawkins has fewer email exchanges and communication targets, our proposed method cannot capture the behavioral similarity between Paris Stamatopoulos and Kevin Hawkins, which will lead to the failure of partner recommendation. For those unrelated individuals, e.g., Alessandro Preziosi and Joel Anair, their communication targets are almost different and thus they do not have email exchanges, resulting in the low behavioral similarity between them, and our method will not recommend each other to them. Therefore, the outstanding performance of our TSSN model in this paper still partly validates its general effectiveness, which could be further proved on more comprehensive datasets in the future.}

\section{Conclusion} \label{sec:Conclusion}
In this paper, we model partner recommendation across different roles as link prediction task in OSS email networks. Particularly, we construct TSSN to retain both temporal and structural information of email network. And then, we propose a random walk embedding method namely TBW, to make recommendation by leveraging embeddings learned from structural properties, temporal information, and individual real identities. Furthermore, we adopt TBW for link prediction on realistic OSS email networks, and compare our method with a number of random walk based embedding methods. Experimental results demonstrate the effectiveness of our method and indicate that TSSN can better capture the temporal information of email networks. For future work, we hope to apply deep learning methods to expand our methods, and utilize OSS unique code repositories to establish social-technical TSSN so as to further improve partner recommendation.

\bibliographystyle{IEEEtran}
\bibliography{Manuscript}

\begin{IEEEbiography}
[{\includegraphics[width=1in,height=1.25in,clip,keepaspectratio]{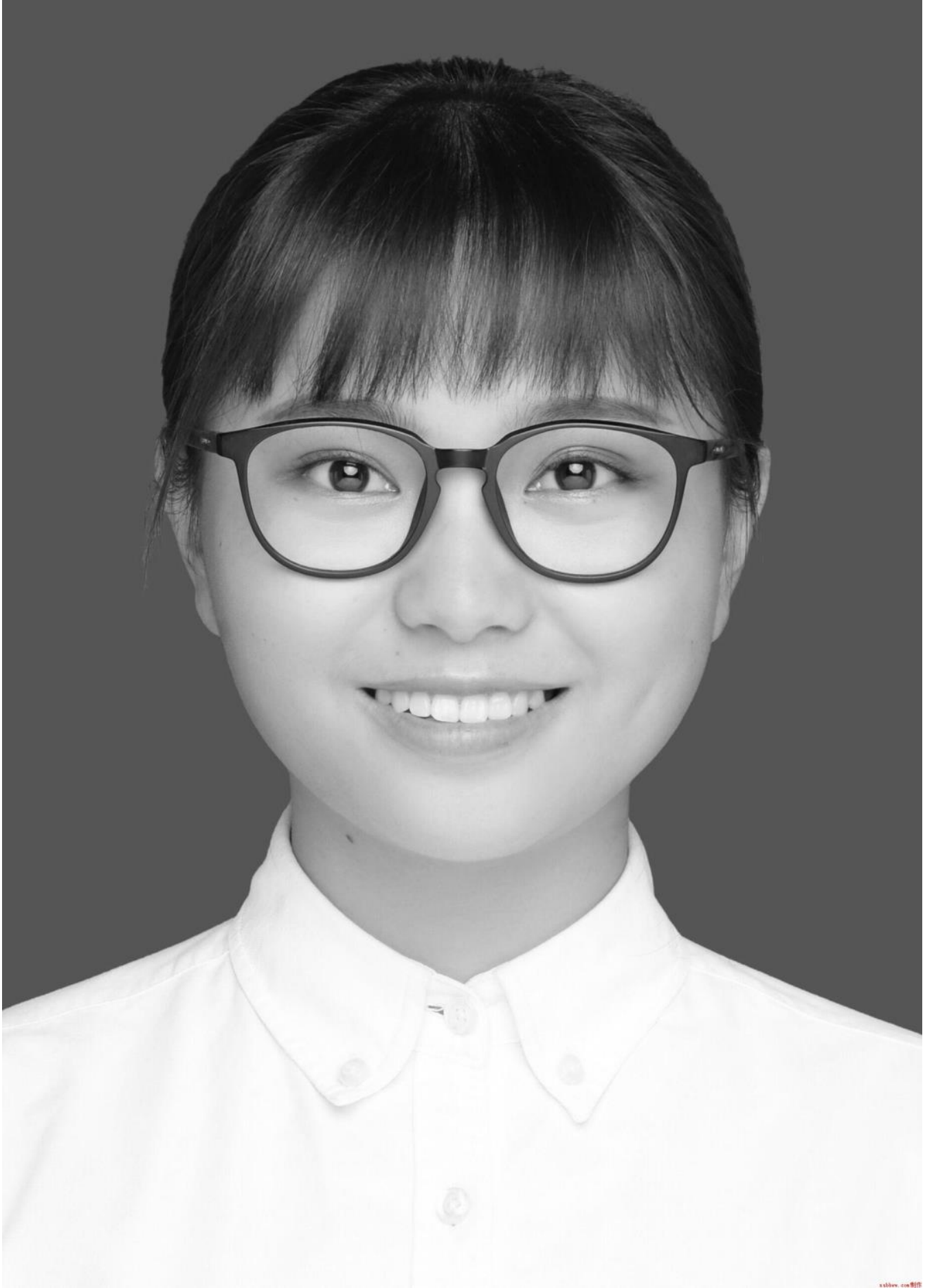}}]{Yunyi Xie}received BS degree from Shaoxing University, Shaoxing, China, in 2019. She is currently pursuing the MS degree at College of Information Engineering, Zhejiang University of Technology, Hangzhou, China. Her current research interests include social network analysis and open source software data research.
\end{IEEEbiography}

\begin{IEEEbiography}
[{\includegraphics[width=1in,height=1.25in,clip,keepaspectratio]{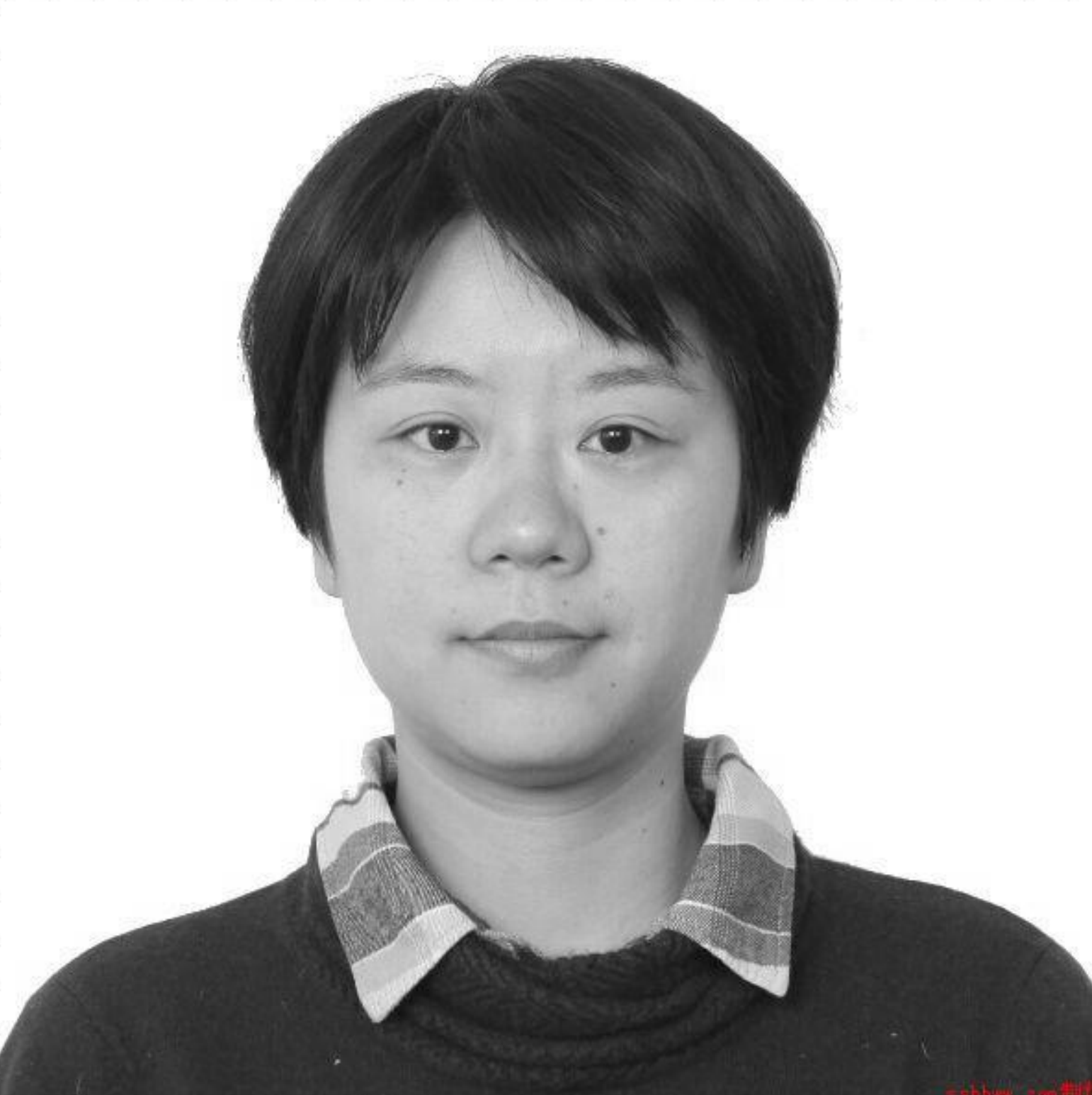}}]{Jinyin Chen}received BS and PhD degrees from Zhejiang University of Technology, Hangzhou, China, in 2004 and 2009, respectively. She studied evolutionary computing in Ashikaga Institute of Technology, Japan in 2005 and 2006. She is currently an associate professor in college of information engineering, Zhejiang University of Technology. Her research interests include evolutionary computing, data mining and deep learning algorithm.
\end{IEEEbiography}

\begin{IEEEbiography}
[{\includegraphics[width=1in,height=1.25in,clip,keepaspectratio]{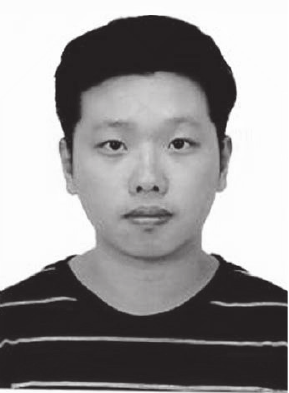}}]{Jian Zhang} received B.S. degree in automation from Zhejiang University of Technology, Hangzhou, China, in 2017. He is currently working toward the doctor' degree in the college of information and engineering, Zhejiang University of Technology, Hangzhou, China. He is currently focusing on network analysis and graph neural networks.
\end{IEEEbiography}

\begin{IEEEbiography}
[{\includegraphics[width=1in,height=1.25in,clip,keepaspectratio]{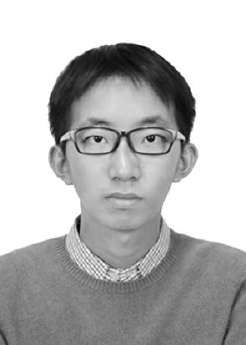}}]{Xincheng Shu} received the B.S. and M.Sc degree from Zhejiang University of Technology, Hangzhou, China, in 2017 and 2020, respectively. He is currently pursing the Ph.D. degree with the control science and engineering, Zhejiang University of Technology. His research interests include information diffusion and machine learning.
\end{IEEEbiography}

\begin{IEEEbiography}
[{\includegraphics[width=1in,height=1.25in,clip,keepaspectratio]{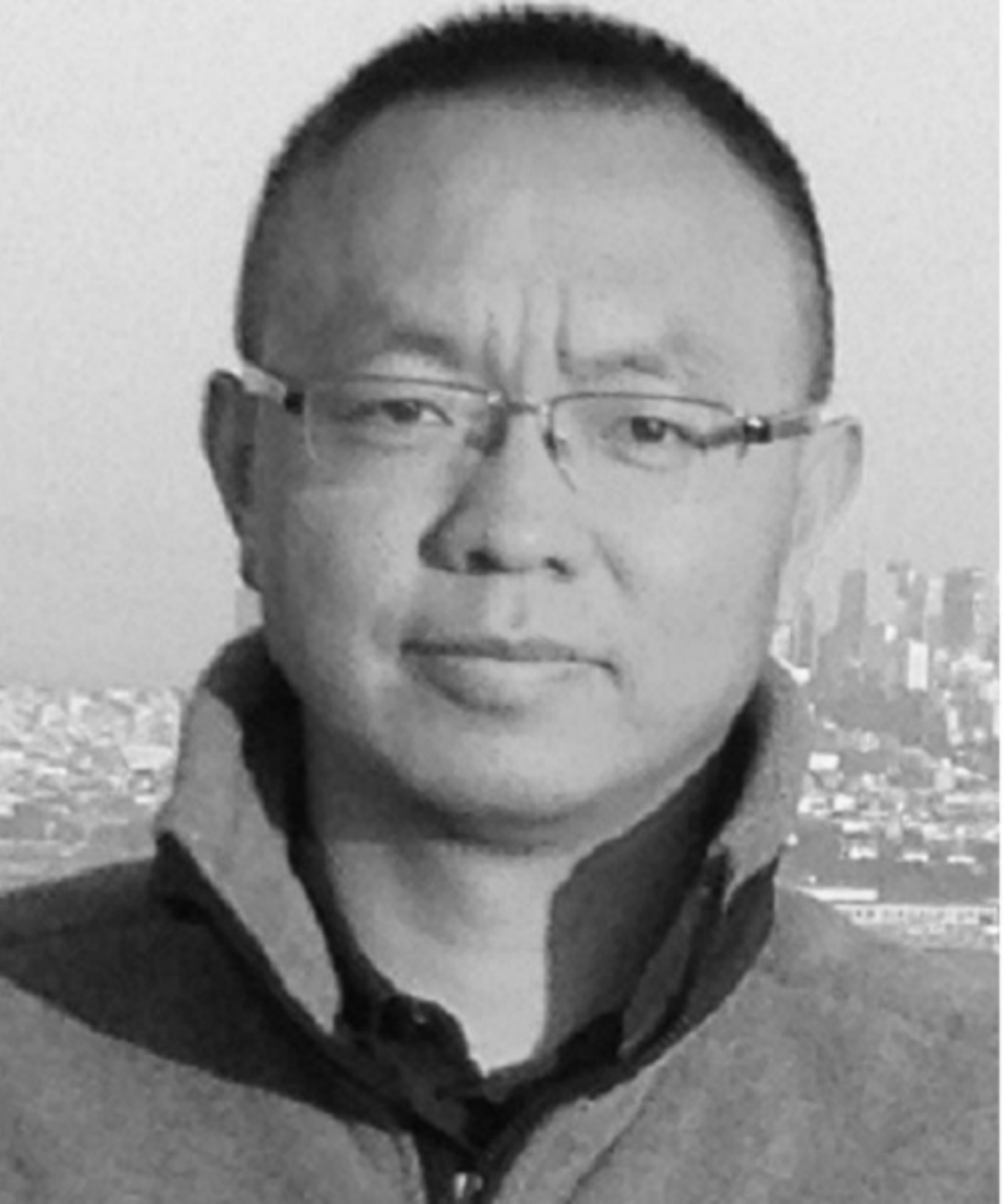}}]{Qi Xuan} (M'18) received the BS and PhD degrees in control theory and engineering from Zhejiang University, Hangzhou, China in 2003 and 2008, respectively. He was a Post-Doctoral Researcher with the Department of Information Science and Electronic Engineering, Zhejiang University, Hangzhou, China from 2008 to 2010, and a Research Assistant with the Department of Electronic Engineering, City University of Hong Kong, Hong Kong SAR, China in 2010 and 2017. From 2012 to 2014, he was a Postdoctoral Fellow with the Department of Computer Science, University of California at Davis, USA. He is currently a Professor with the College of Information Engineering, Zhejiang University of Technology. His current research interests include network-based algorithms design, social networks, data mining, cyberspace security, machine learning, and computer vision.
\end{IEEEbiography}

\end{document}